# Colwell's Castle Defence: A Custom Game Using Dynamic Difficulty Adjustment to Increase Player Enjoyment

Anthony M. Colwell and Frank G. Glavin

College of Engineering and Informatics,
National University of Ireland, Galway, Ireland.
a.colwell1@nuigalway.ie, frank.glavin@nuigalway.ie

**Abstract**. Dynamic Difficulty Adjustment (DDA) is a mechanism used in video games that automatically tailors the individual gaming experience to match an appropriate difficulty setting. This is generally achieved by removing pre-defined difficulty tiers such as *Easy*, *Medium* and *Hard*; and instead concentrates on balancing the gameplay to match the challenge to the individual's abilities. The work presented in this paper examines the implementation of DDA in a custom survival game developed by the author, namely Colwell's Castle Defence. The premise of this arcade-style game is to defend a castle from hordes of oncoming enemies. The AI system that we developed adjusts the enemy spawn rate based on the current performance of the player. Specifically, we read the Player Health and Gate Health at the end of each level and then assign the player with an appropriate difficulty tier for the proceeding level. We tested the impact of our technique on thirty human players and concluded, based on questionnaire feedback, that enabling the technique led to more enjoyable gameplay.

## 1 Introduction

Maintaining a player's interest and engagement for long periods of time is a challenging problem faced by all game developers. This paper will attempt to address this problem by creating a custom game and implementing a mechanism to detect the players' abilities and best match the difficulty to result in a more enjoyable gaming experience. One of the key engagement factors, when it comes to gaming, is the difficulty level. If a game is too easy it quickly becomes boring but if a game is too hard then players can become frustrated and stop playing [1]. Individual players will have different qualities in terms of hand-eye coordination, reflex speeds, and their own personal playing preferences. It is very difficult for game developers to find a good balance which encompasses all the attributes of a gamer and place them into a simple static difficulty rating system. This simple system is frequently used, however, if a player is in-between the skill levels of, for example medium and hard, it would result in neither being a good option, leading to the game becoming frustrating or boring [2].

### 1.1 The Research Problem and Overall Perspective

The basis for this research originated from playing '*Junkenstein's Revenge*', an arcade-like sub game within the greater game of '*Overwatch*' by Blizzard Entertainment [3]. Our experience with 'Junkenstein's Revenge' was that the medium difficulty level was too easy whereas the hard difficulty level was impossible to succeed at. We grew continually frustrated from only having two options: win relatively easily or almost certainly be defeated.

For this research, we have designed and developed a similar style game and we have introduced a DDA mechanism into the gameplay. This eliminates the standard easy, medium and hard settings with the difficulty level being dynamically determined based on the current performance of the player. This results in a new gaming experience each time that will challenge players of different skill levels. We believe that the use of DDA in real-time gaming can enhance player enjoyment. In order to test this hypothesis, we created a custom survival game, Colwell's Castle Defence, and carried out multiple gameplay testing sessions with thirty human players in order to compare player enjoyment of the game when DDA is both activated and deactivated.

**1.2 Dynamic Difficulty Adjustment**

Dynamic Difficulty Adjustment (DDA), also known as Auto Dynamic Difficulty (ADD) and Dynamic Game Balancing (DGB), is a technique whereby the game has the ability to alter its difficulty in real-time, based on the current player's abilities [4]. One reason an adaptive technique like DDA is not often used in games is that it can be expensive. When implemented correctly, DDA can be very beneficial to the overall gaming experience, as it offers a new experience each time it is played, removing the obvious patterns seen in basic scripted games which players can use to fast track their progress.

The gaming market has become one of the largest entertainment markets in the world in recent years with it reaching just over the 100 Billion mark in 2016 [6]. Despite this continual growth, there is often a dissatisfaction amongst gamers. There is a sense that a rigorous difficulty allocation does not have the capabilities to provide an optimal challenge in games to accommodate individual player's attributes [7]. This is backed up by Koster [8] who states the static difficulty tiers, that players must choose from, are inadequate as players may not be able to access a challenge that best matches their skill set. In a bid to tackle this issue there has been a growth in the number of studies conducted to identify if DDA can resolve this issue by offering players a game that is tailored to match their abilities [9].

**1.3 Flow**

*Flow* is a concept introduced by Mihaly Csikszentmihalyi [10], a professor of psychology, to explain happiness. Since introducing the concept of flow, it has become a fundamental concept in the field of positive psychology in which Csikszentmihalyi has gone on to write several books related to all areas of life. It is commonly used in relation to feeling complete focus in an activity with a high level of enjoyment and fulfilment. This is applicable to many aspects of life, but is also applicable to video games. An example of being in the flow that most gamers could relate to is when one becomes so engrossed in a game, they seem to forget about everything else such as eating or sleeping.

The concept of flow is to keep players within the *flow zone* by ensuring the game gets neither too hard (leading to anxiety), nor too easy (leading to boredom). Since all players' abilities vary, this can be difficult and this is where we believe DDA can be a useful technique for keeping players within this zone. This is in essence, what will be attempted in the game, Colwell's Castle Defence, as a player may be performing poorly or strongly, in which case the DDA will help to guide the player back into the flow zone by using slight adjustments in the difficulty of the game [11].

## 2      Related Research

Jenova Chen deigned his own game called *FlOw* [12]. This game was developed to test his theory that allowing players to make their own choice within the game to alter the difficulty can improve the enjoyment they experience. The game is purposely simplistic in which the player moves an organism around the screen with the use of the mouse, with the object being to eat other organisms to grow and advance levels to meet and attempt to consume larger organisms. The choice in difficulty comes by the player having control of avoiding the larger organisms and consuming smaller ones until they feel they are more equipped to engage these. As a result, players play the game at their own pace and subject to the player, can either engage in a slower paced game or an intense fast paced game. Within two weeks of Chen releasing the game it had almost 350,000 downloads and was awarded the Game of the Month on EDGE magazine in May 2006 [11].

   Another study that used the concept of flow as a core component was completed by Hunicke and Chapman [13] in which they used a tool, *Hamlet*, in the game 'Half Life' by Valve [14]. The Hamlet architecture was used to read data of the players game and make adjustments when required. They use the concept of flow to maintain players in engaging interactions and therefore within a state of flow. This was achieved through reacting to events by altering parameters such as damage and inventory. Hamlet was used to adjust *reactively* (adjusting in-game parameters such as accuracy and damage of Non-Player Characters (NPC) attacks) and *proactively* (adjusting NPCs health and damage before they are spawned into the game) in an attempt to stop players entering into loops whereby they would die repeatedly or struggle in game, thus leading to frustration. Their aim was to discover if it was possible to implement a system that could make adjustments seamlessly but were unclear as to the results of the study. The authors mentioned that they intend to test on participants wearing heart rate monitors to further the study.

   A study was carried out by Lankveld et al. [15] in which they proposed a model to balance a game using the *incongruity theory* as the basis of their study. Incongruity is the difference of internal and external complexity which produces a negative or positive incongruity. This is similar to the concept of flow in which they used the model of incongruity to determine the enjoyment of the participant. If the game becomes too complex (difficult) the user shows a positive incongruity value which would suggest the game has become too hard resulting in frustration. The opposite is experienced if a negative value is obtained which means the player deems the game too easy and leads to boredom. The purpose of the study was to determine if they could engage players by maintaining them in an incongruity value as close to zero as possible by dynamically adjusting the difficulty of the game. Therefore, ensuring that they are being challenged throughout their experience and remaining within a state of flow, thus improving enjoyment.

## 3      Methodology

We designed and developed a custom arcade-style castle defender game using the *Unity* software development environment [16], in order to test our hypothesis. In the interest of brevity, we have summarised the most important points below:
- It is an arcade-styled game with a top-down camera that follows the playable character around a small enclosed area.
- The playable character has two swords as weapons with two main attack functions, a chop action and a spinning attack option.

- There are two types of enemies, *Tankers* and *Zombies* which both spawn into the game based on a random ratio value.
- The Tankers traverse towards the playable character, attacking when in range.
- The Zombies ignore the playable character and instead traverse towards the defence object (the castle Gate) and try to destroy it.
- Both the playable character and the Gate can sustain 10 hits before being destroyed, which then ends the game.
- The playable character regenerates full health at the end of each level (when all enemies have been killed) whereas the Gate Health does not.
- The game starts with 10 enemies and after all are killed, a new level begins with the base figure of three additional enemies per level.

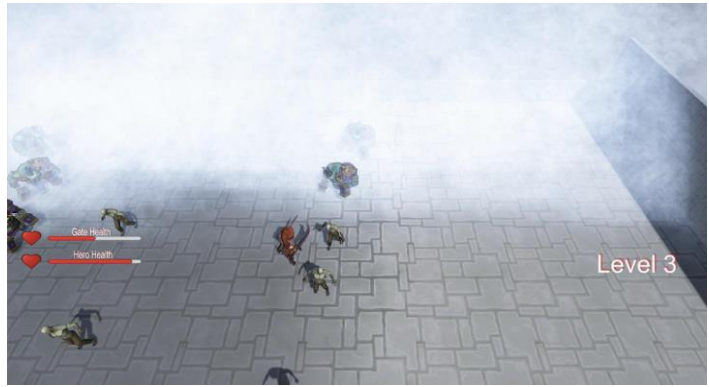

*Figure 1. Image of Colwell's Castle Defence where the playable character and enemies can be seen.*

Once the base game was created, an AI system was then developed in which players are allocated to skill tiers depending on their progress in the game, essentially matching the ability of the player to the difficulty of the game. This was achieved by using the Health values of both the gate object and the playable character, in order to allocate the player to a tier at the end of each level of gameplay. Once this was allocated, a new level would begin with a new number of enemies equal to the skill tier of the player. Enemies were determined to have the largest impact on the difficulty of the game, while any alterations made remained difficult for players to notice. Parameters such as speed and damage were experimented with but they were deemed to be too noticeable to human players during initial tests. As a survival game, enemies per wave was a value that was always going to increase as this genre of game always increases the challenge for each consecutive level. We incorporated DDA to vary the increase in difficulty based on player performance. The additional number of enemies per wave, based on the player's tier allocation can be seen in Table 1 below.

*Table 1. Table displaying the spawn increase per level based on the Tier allocation of a player.*

| Performance | Tier | Spawn Increase |
|---|---|---|
| *Poor* | T1 | + 1 Enemy |
| *Below Average* | T2 | + 2 Enemies |
| *Average* | T3 | + 3 Enemies |
| *Above Average* | T4 | + 4 Enemies |
| *Very Good* | T5 | + 5 Enemies |

The calculation used to determine the tier allocation for players is based on the average of the Gate and Player Health as follows:

$$\frac{\text{Gate Health (GH)} + \text{Player Health (PH)}}{2}$$

Since the lowest health achievable at the end of a round is 10 (receiving another hit would result in game ending), the lowest result that this calculation can achieve is 10 ((10+10)/2) and the largest value is 100 ((100+100)/2). These figures were used as the maximum and minimum in the creation of the tiers. The difference was then divided out among the five tiers to give a range which can be seen in Table 2 below.

*Table 2. The ranges used to determine the Tier allocation of a player.*

| Performance | Tier | Range |
|---|---|---|
| *Poor* | T1 | 10 – 27 |
| *Below Average* | T2 | 28 – 45 |
| *Average* | T3 | 46 – 63 |
| *Above Average* | T4 | 64 – 81 |
| *Very Good* | T5 | 82 - 100 |

**3.1 Testing**

There were four preliminary rounds of testing carried out before the final full test phase. The first two phases were used to balance the gameplay of Colwell's Castle Defence, to ensure all players reached a minimum of Level 3 and a maximum of around Level 8-10. The reason for this was to ensure players did not die too early, resulting in frustration for the game being overwhelming and that players did not reach too high a score, therefore becoming bored of the lack of challenge posed to them. It is important to note that for consistency, the same 6-8 participants with varying gaming backgrounds and abilities were used throughout the first four phases of testing to ensure accurate feedback was obtained (an initial eight participants took part in Phase 1 and Phase 2 of testing but only six of the original group were available for Phase 3 and Phase 4).

After the first round of testing there were several changes made to the game, such as the character's speed decreased along with changes to the enemies such as increased range to give a better balance to the game. The most significant change made was to get rid of the Gates ability to regenerate health between levels as it proved too easy to defend it during this initial phase of testing. These alterations were made based on the feedback received from the participants.

We collected the following player data: gaming background, levels reached, time taken to play a game, the number of enemies remaining upon the games end and the difficulty perceived by the player. This data was used to determine how the game could be improved and what were the problem areas that needed to be addressed. The main issue encountered was stronger players feeling the early levels were too easy, leading to boredom. For the weaker players, it was how quickly the game was ending and not being able to progress through many levels which led to frustration. The weaker players, however, reported more enjoyment experienced than the stronger players. It was for these reasons alterations were made to the game in an attempt to give it more balance. The average results of all participants (using three *Weak* and *Strong* players with two *Average* players as a base for testing) of the first phase of testing can be seen below in Table 3.

*Table 3. Data collected from the eight participants of Phase 1 of testing, highlighting the difference in time taken, level reached, difficulty and enjoyment perceived for Colwell's Castle Defence.*

| Gaming Background | Time Taken | Level Achieved | Difficulty (1-5) | Enjoyment (1-10) |
|---|---|---|---|---|
| **Weak** | 5 minutes | 3.5 | 3.5 | 7 |
| **Medium** | 6 minutes | 5 | 3.5 | 7.5 |
| **Strong** | 7 minutes | 8 | 3 | 6 |

The second phase of testing was used to determine if the changes made to the game were of benefit and after having the same participants replay the game, positive feedback was received and thus the game was deemed to be balanced.

Phase three of testing was used to determine if the DDA system was working as intended. For this, the same participants were again asked to test the game, this time noting the number of additional enemies per wave and the Health values that the player had at the end of each level. This was to ensure that the stronger players were getting a challenge earlier in the game whilst the weaker players were eased into it, ensuring higher scores and longer game time. During this testing phase, it was noted that the stronger players were in fact engaged in a competitive sense earlier in the game due to the higher spawn rates with many receiving five additional enemies for the first 2-3 Levels compared to the three of the base game, meaning shorter games and a more challenging experience. However, for the weaker players there was very little difference noted with many still receiving the three additional enemies, instead of the 1-2 that was intended. This was due to the Gate still maintaining high health in the early rounds, resulting in Tier 3 allocations despite the low playable character health. This meant the DDA system needed some adjustment to counter this but also to have no impact on how it performs on the stronger players.

We addressed this problem by altering the Tier allocation calculation, by essentially halving the relevance of the Gate Health. This meant strong players still maintained their high Tier allocation but the weaker players who sustained large damage to the playable character in the earlier rounds got placed into a lower Tier allocation, thus reducing the pace of the increased difficulty. The new calculation is as follows:

$$\frac{(\text{Gate Health} / 2) + \text{Player Health}}{2}$$

With this new calculation came the need to adjust the ranges of the Tiers themselves. There was no change to Player Health and the amendments made to the Gate Health had the following impact on the ranges used in the Tier allocation which can be seen in Table 4.

*Table 4. The difference in ranges used to determine the Tier allocation after phase 3 of testing*

| Performance | Tier | Old Range | New Range Used |
|---|---|---|---|
| *Poor* | T1 | 10 – 27 | 7.5 – 21 |
| *Below Average* | T2 | 28 – 45 | 21.5 - 35 |
| *Average* | T3 | 46 – 63 | 35.5 – 49 |
| *Above Average* | T4 | 64 – 81 | 49.5 - 62 |
| *Very Good* | T5 | 82 - 100 | 62.5 - 75 |

Once these new values were implemented into the game, they needed to be retested which meant asking the same participants to again play the game. This was phase four of testing, which showed promising results as it did not affect the Tier allocation of the stronger players but did place the weaker players in Tier 2 from the outset, meaning they were eased into the game for Level 1-2 and then placed in Tier 1 as the Gate Health reduced, meaning they reached higher Levels and played for longer. After these four phases of testing, it was felt that the game was sufficiently balanced both in terms of gameplay and the DDA system. From here, it was time to move onto the full testing, phase five, which we present and discuss in the following section.

## 4      Results

In order to ensure that our DDA implementation was working as desired, we selected a diverse test group with varying gaming backgrounds for our experimentation. We asked the participants in the survey during testing to rate their gaming experience from 1-5 (weak to strong). The results displayed the diverse range of skills that were required with 20% answering a Level of 3 out of 5 with an almost 50/50 split either side of average.

The main factors that were analysed to back up any change in enjoyment experienced by participants were Levels reached, time played and difficulty perceived. If these areas showed a change in the same direction as the overall enjoyment itself, it would help to prove/disprove our hypothesis.

Thirty participants took part in the experimentation, with these being split in to *Weak*, *Average* and *Strong* players to determine the desired adaption to their gameplay (easier for weak players and more difficult for strong players). The participants were brought into a test room in small groups and asked to play each game three times(without DDA and then with DDA), not knowing which they were playing and the desired results of the testing. It was decided to test the game without DDA first as a basis of results meaning those of the game with DDA were accurate. The argument for players learning from the first game are well founded, but only effect the *Weak* players, as it is desired for *Strong* players to achieve lower levels in the game with DDA. We chose the eight weakest and eight strongest players based on their gaming backgrounds and their results achieved from our analysis, as the average players were already performing within a state of flow, and the game was considered balanced with the DDA mechanism doing little to change their experience. The comparison tables were all made using the same eight weak and strong participants.

### 4.1      Average Levels Reached

The average levels reached by the participants are a strong determinant when the overall enjoyment is considered. As one of the main issues raised during testing was the stronger players finding it boring in early levels and the weaker players dying early, the DDA needs to address this by enabling weaker players to progress further in the game and by making it harder for the stronger players.

As can be seen from Figure 2 below, the average levels of the weaker players were increased across the board. This would suggest that they were engaged for longer and held greater satisfaction in their performance in contrast to playing without DDA enabled. As can be seen from Figure 3 below, strong players reached lower levels in each case. This would suggest the earlier levels were harder for them introducing more enemies which meant a more challenging experience was obtained earlier

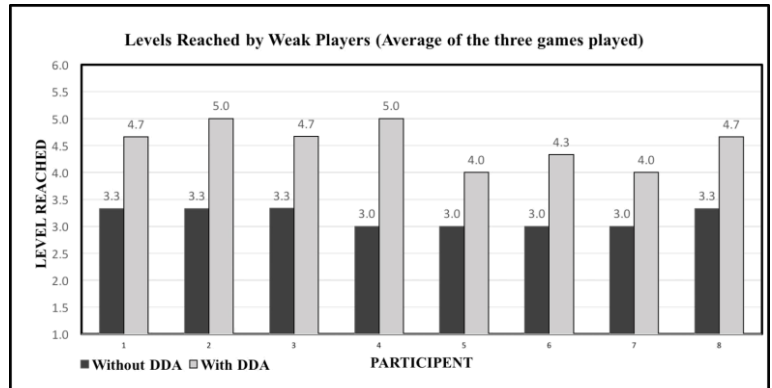

*Figure 2. The Levels reached during testing of both games for the weak players.*

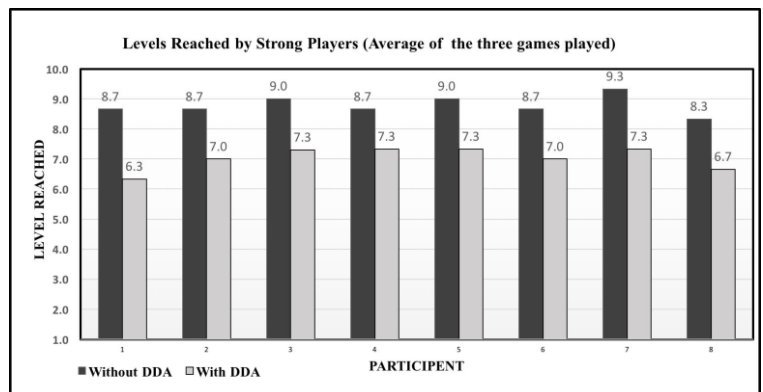

*Figure 3. The Levels reached during testing of both games for the strong players.*

### 4.2  Average Time Taken per Game

The next determinant in overall enjoyment looked at was the average time taken to play a game for both the weak and strong players. The sense of boredom was experienced from the stronger players due to the lack of challenge experience and therefore a reduction in time played would suggest more competitive gaming experience. For the weaker players it is the opposite, in essence, as they felt frustrated at the lack of time playing and progress made and therefore, an increase in time taken would be beneficial to their gaming experience.

From Figure 4 and 5 below we can see that the weaker players experienced longer games, coupled with the higher levels reached. For stronger players, the objective of DDA is to show a reduction of time taken to play a game. We believe that this helps remove a sense of boredom experienced and enables players to play the game repeatedly, as a game taking too long would deter people from repetitive play and possibly from the game altogether. This coupled with lower levels achieved could suggest a greater enjoyment experienced by the players. This is confirmed from our survey analysis described later.

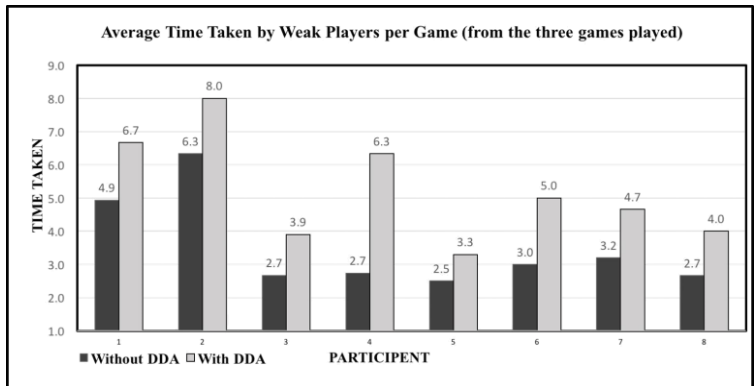
*Figure 4. The average time taken for both games during testing of the weak players.*

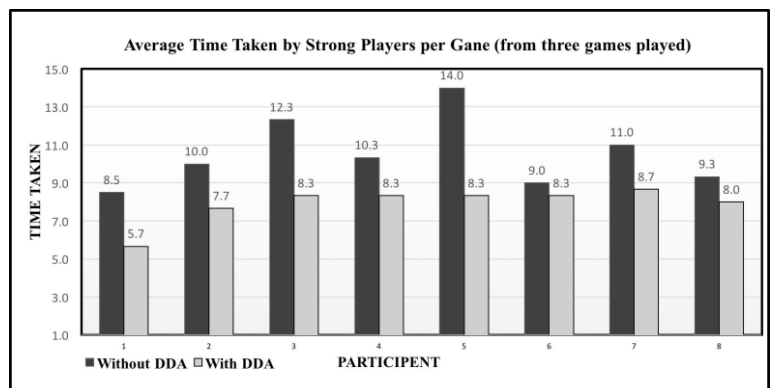
*Figure 5. The average time taken for both games during testing of the strong players.*

### 4.3 Perceived Difficulty

As flow is measured in challenge versus ability, the difficulty felt by the participants is arguably the largest determinant in the enjoyment experienced. In order to maintain a player within a state of flow, the game must not become too easy or too hard to avoid boredom or frustration. For the weaker players, they experienced frustration with the game being too difficult and ending early. For the stronger players, they experienced boredom as they felt the game was too easy. Therefore, the DDA system needs to address this by bringing down the difficulty for weaker players and making it harder for the stronger players.

We can see from Figure 6 and 7 that this was achieved as, in the majority of cases, the weaker players showed a reduction in perceived difficulty, suggesting they found the game with DDA to be more balanced. Likewise, for the stronger players finding the game with DDA implemented to be more balanced, but for them it shows an increase in difficulty. These figures can be further backed up by our survey analysis. Having asked the participants for a 1-5 rating of difficulty of both games, the game with DDA enabled proved far more balanced with 63% of participants answering so, compared to the 33% that found the game with no DDA to be balanced.

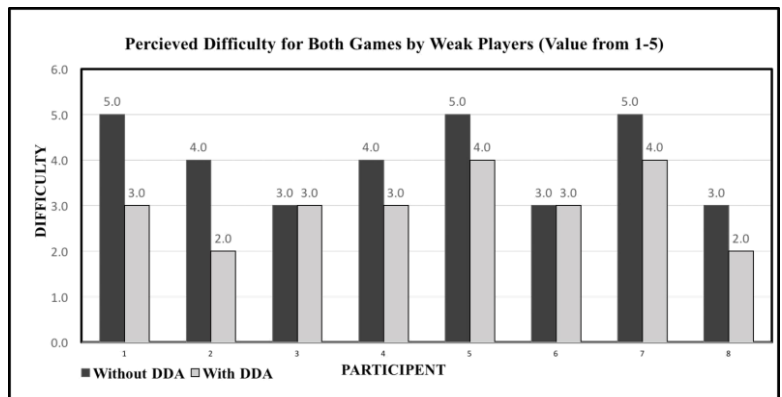
*Figure 6. The perceived difficulty of both games allocated by the weak players.*

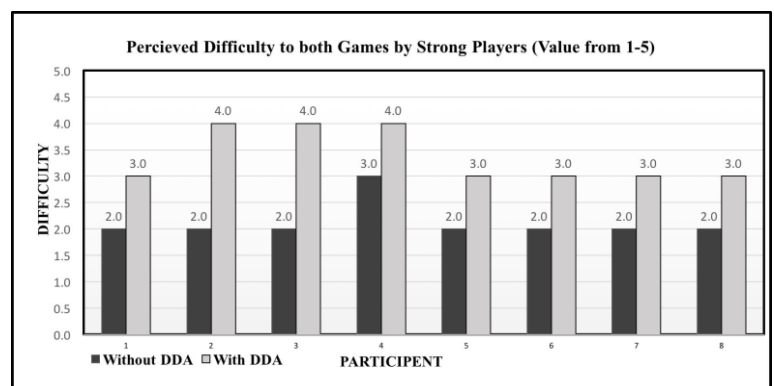
*Figure 7. The perceived difficulty of both games allocated by the strong players.*

### 4.4    Enjoyment

All results shown is this section are contributory factors to the main aim of this study, to improve player enjoyment. Each of these factors contain positive results that build into the overall enjoyment but it is not a guarantee that players will prefer the game with the inclusion of the AI system. For the weaker players, the challenge was to ensure higher levels reached, longer game time and a lower difficulty in a bid to improve overall enjoyment. Due to the positive results across all of these factors, we can see in Figure 8 that this did in fact improve the enjoyment experienced by these participants with the average perceived enjoyment increasing from a 5.8 to an 8 (out of 10).

For the stronger players, the aim was to reduce the levels reached, reduce the time spent playing and to increase the difficulty in a bid to improve player enjoyment. We can see from Figure 9, due to the success of the previous factors that the overall perceived enjoyment was increased from an average of 5.9 to 8.3, an increase of 40.7 percent.

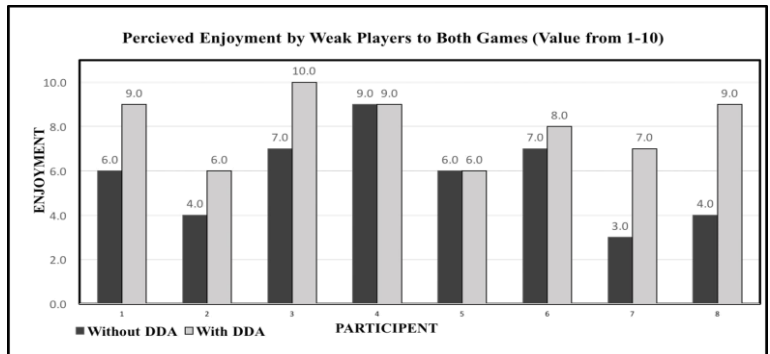

*Figure 8. The perceived enjoyment allocated to both games by the weak players*

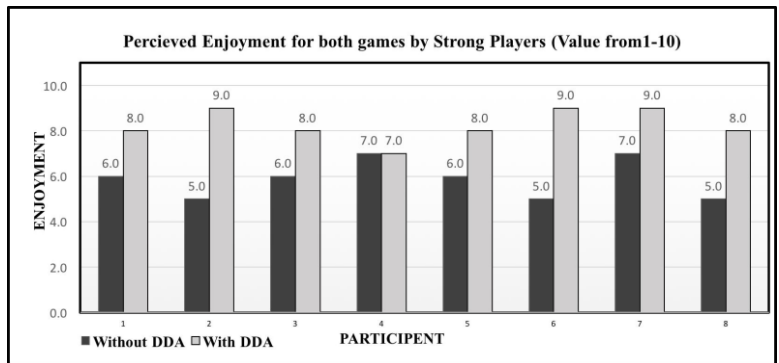

*Figure 9. The perceived enjoyment allocated to both games by the strong players.*

# 5 Conclusion

The results gathered from this study were all positive in varying degrees. The main hypothesis was to determine if the inclusion of a DDA system can improve a player's perceived enjoyment in a game. To ensure the data was accurately analysed to answer the hypothesis, the other components of enjoyment must also be examined to ensure these results are accurate. For this, *Levels Reached*, *Time Taken* and *Perceived Difficulty* were also analysed to show how they were affected by the balancing mechanism. An overview of these results can be seen in Table 5 below.

*Table 5. Table highlighting a summary of all the results analysed in the previous section.*

| | Averaged Results Overview | | |
|---|---|---|---|
| | **Without AI** | **With AI** | **Difference** |
| **Level Reached** | Weak: 3.2 <br> Strong: 8.8 | Weak: 4.5 <br> Strong: 7 | Weak: + 1.3 <br> Strong: - 1.8 |
| **Time Taken** | Weak: 3.5 <br> Strong: 10.6 | Weak: 5.2 <br> Strong: 7.9 | Weak: + 1.7 <br> Strong: - 2.7 |
| **Difficulty** | Weak: 4 <br> Strong: 2.1 | Weak: 3 <br> Strong: 3.4 | Weak: - 1 <br> Strong: + 1.3 |
| **Enjoyment** | Weak: 5.8 <br> Strong: 5.9 | Weak: 8 <br> Strong: 8.3 | Weak: + 2.2 <br> Strong: + 2.4 |

From the results shown in Table 5, it can be seen that all the areas analysed showed that the use of our proposed technique led to positive trends in overall enjoyment. The results show an increase of 2.2 (out of 10) in perceived enjoyment by the weak players, showing an increase of 37.9%. There was an increase of 2.4 (out of 10) in perceived enjoyment by the strong players, which is an increase of 40.7% in overall enjoyment experienced. Therefore, the inclusion of a simple enemy spawn-based DDA system into our custom game can improve the enjoyment experienced by those who play it.

## 6  Future Work

Firstly, a more complex game with more components and required strategies could lead to more interesting insights about the player feedback. This could include additional enemies, health pick-ups and special abilities. Additional DDA adjustments, such as the speed and damage of NPCs, mobility of the player character and rate of damage for the Gate would be worthwhile to investigate in more detail. Each of these areas detailed above provide the opportunity for further research and study to be carried in the field of Dynamic Difficulty Adjustment for improving player enjoyment in video games.